# Balanced incomplete block designs and exact satisfiability

Bernd R. Schuh

Dr. Bernd Schuh, D-50968 Köln, Germany; bernd.schuh@netcologne.de

**Abstract.**
The paper explores the correspondence between balanced incomplete block designs (BIBD) and certain linear CNF formulas by identifying the points of a block design with the clauses of the Boolean formula and blocks with Boolean variables. Parallel classes in BIBDs correspond to XSAT solutions in the corresponding formula. This correspondence allows for transfers of results from one field to the other. As a new result we deduce from known satisfiability theorems that the problem of finding a parallel class in a partially balanced incomplete block design is NP-complete.

**Introduction.**

In the theory of block designs one is mainly concerned with the *existence* and *construction* of designs with certain structural properties. Whereas in the investigation of CNF formulas (structural) existence is generally assumed and necessary and/or sufficient conditions for various types of (logical) *satisfiability* are investigated. That may be the reason why similarities between both fields have not been explored extensively, if at all. There are questions of satisfiability, however, which strongly depend on structural properties of the Boolean formula. One example is exact satisfiability (XSAT), i.e. the fulfillment of a conjunction of clauses with no more than one true variable in each clause. XSAT restricted to regular linear CNF formulas can be described as a purely structural problem. Also in block designs similar structural properties, which bear some resemblance to the satisfiability questions in Boolean systems, are known and have been the subject of investigations, namely the notion of parallel classes and the property of resolvability. Quite generally we believe that both fields can benefit from each other's results. As one example we take the effect of regularity on exact linear formulas. Regularity forces uniformity in such formulas, a fact that apparently has long gone unnoticed, although it was well known and used in the corresponding field of balanced incomplete block designs. As a second example we apply investigations on the complexity of XSAT restricted to linear CNF formulas (see [1,2,3]) to the decision problem of finding a parallel class in a partially balanced incomplete block design. We find that this decision problem is NP-complete. Thus we generalize an earlier result which states NP-completeness for Steiner triple systems only, [4].
The paper starts with basic definitions, continues with setting up the correspondence between BIBDs and regular XLCNF as well as between the notions of a parallel class in BIBDs and exact satisfiability in propositional logic. It then gives some more or less trivial examples of transfers between the two fields, mainly for illustrational purpose. Readers, who are familiar with both fields and interested in new results only are advised to go to definition 6 and the following theorem directly.

**Basic definitions**.
    a. **Boolean CNF formulas**.
Throughout this communication I will adopt the notation for Boolean formulas used in [1,2,3], slightly generalized to make contact to a larger class of block designs. A Boolean formula in *conjunctive normal form* (CNF) by definition is a conjunction of clauses, where each clause is a disjunction of literals. A *literal* is an occurrence of a Boolean variable or its negation. We assume throughout the paper that all formulas are *monotone*, i.e. all literals are of one kind (either variable or its negation). In a *linear* CNF formula any two clauses have at most one variable in common. The class of such formulas is denoted by LCNF. One further restriction is *exact linearity. In such* systems any two clauses have <u>exactly</u> one variable in common (XLCNF). As a generalization we define

*Definition 1*:
An exact $\lambda$-connected CNF formula is a set $F$ of clauses $C_i$ over a variable set $Var(F)$ with

(i)      $C_i \in 2^{Var(F)}$

(ii)      any two clauses have exactly $\lambda$ variables in common.

We write this class of formulas as $X^\lambda CNF$.

Thus XLCNF corresponds to $\lambda = 1$, whereas LCNF consists of conjunctions of elements from $X^1$- and $X^0$-CNF.
A more general definition is a $\lambda$-*connected CNF formula* (without the "exact") as before with condition (ii)) replaced by
(ii')      any two clauses have <u>at most</u> $\lambda$ variables in common.
For those formulas we write $^\lambda CNF$.

An additional structural restriction is *l-regularity*, the property that each variable occurs exactly $l$ times. It is denoted by a superscript $l$, e.g. CNF$^l$. Finally, *uniformity* means that every clause contains the same number of literals. If this number is $k$, one writes e.g. $k$-CNF.
If not stated otherwise the following notation is used for a CNF formula $F$: $m = |F|$ for the number of clauses of $F$, $n = |Var(F)|$ for the number of variables of $F$, $l(a) = |\{C \in F : a \in Var(C)\}|$ for the occurrence of variable $a$, $k_C = |Var(C)|$ for the number of variables in clause $C$.

Each monotone CNF formula is structurally defined by its incidence matrix $F \triangleq (f_{ij})$ where $i \in \mathbb{N}_m$ runs through all clauses, $j \in \mathbb{N}_n$ through all variables of $F$. We use the same symbol $F$ in the following for the incidence matrix and the Boolean formula, since the context clarifies what is meant.

*Definition 2* (*Incidence matrix* of Boolean formula in CNF form):
$f_{ij} = 1$ iff variable $a_j \in Var(C_i)$, $f_{ij} = 0$ otherwise; $i \in \mathbb{N}_m$ and $j \in \mathbb{N}_n$.

    b.      **block designs**.
According to classic design theory one uses:

*Definition 3*
A balanced incomplete block design (BIBD) is a pair of finite sets $(X, B)$ of $m = |X|$ *points* and $n = |B|$ *blocks*, also called lines, with the following properties:
(i)      Each block is a true subset of $X$
(ii)      every block contains exactly $l < m$ points
(iii)      every pair of points is contained in exactly $\lambda$ blocks,
see e.g. [5].
(I have adjusted the notation in order to make later comparison simpler).

The condition $l < m$ makes a block design *incomplete*, the exactness in condition (iii) gives rise to the term *balanced*. If this exactness is relaxed to
   (iii')   every pair of points is contained in <u>at most</u> $\lambda$ blocks
one speaks of a *partially balanced* incomplete block design or *partial BIBD*.

A consequence of the defining properties is that every element of $X$ is contained in the same number, say $k$, of blocks. This property is sometimes used as an additional defining property. But that is unnecessary, the three parameters $(m, l, \lambda)$ are sufficient to define a BIBD.
$k$ can be calculated from $m$, $l$ and $\lambda$ via
(BD1)   $\lambda(m-1) = k(l-1)$
Also the number of blocks $n$ in a BIBD is given by the defining parameters via:
(BD2)   $nl = mk$
One necessary condition for the existence of a $(m, l, \lambda)$-BIBD, of course, is that $k$ and $n$ are natural numbers.
An existing BIBD can be characterized uniquely by an $(m, n)$-matrix $G$, called incidence matrix:

*Definition 4* (incidence matrix of a block design)
$g_{ij} = 1$ if point $i$ of $X$ is contained in block $j$ of $B$, and $g_{ij} = 0$ otherwise.

**Correspondence between BIBDs and XLCNF formulas.**
Consider an $l$-regular $^\lambda CNF$ with $n$ variables. In the standard view one defines clauses as subsets of variables $C_i = \{a_{j_1}, a_{j_2}, ..., a_{j_{k_i}}\}$. Since each variable $a$ belongs to $l(a)$ clauses, one can set up a correspondence between variables and the subset of clauses to which each variable belongs, given by:

$$(COR1) \quad \begin{aligned} &\Phi : Var(F) \to 2^F \\ &\Phi(a_j) = \{C_i : a_j \in Var(C_i), i \in \mathbb{N}_m\} \end{aligned}$$

Similarly in a partial $(m, l, \lambda)$-BIBD blocks can be regrouped according to the points they contain via the following mapping:

$$(COR2) \quad \begin{aligned} &\Psi : X \to 2^B \\ &\Psi(p_i) = \{b_j \in B : p_i \in b_j, j \in \mathbb{N}_n\} \end{aligned}$$

With these settings we can state:

*Theorem*: (correspondence theorem)
The set theoretic structure of a $l$-regular $\lambda$-connected CNF formula $F$ with $m = |F|$ and a variable set $Var(F)$ defines a partial $(m, l, \lambda)$-BIBD with $X = F = \{C_1, C_2, ..., C_m\}$ and $B = \{\Phi(a_j) : j \in \mathbb{N}_n\}$. Likewise defines a partial $(m, l, \lambda)$-BIBD with point set $X$ and block set $B$ the structure of an $l$-regular $^\lambda CNF$ with $F = \{\Psi(p_i) : i \in \mathbb{N}_m\}$ and $Var(F) = B$.

*Proof:*
Let the CNF formula be given. According to the definition of a $(m, l, \lambda)$-BIBD we must show that conditions (i) to (iii') are fulfilled. Conditions (i) is valid by definition of the points and blocks. Since in an $l$-regular $^\lambda CNF$ all sets $\Phi(a_j)$ have the same cardinality $l$ condition (ii) is valid. Finally, in a $^\lambda CNF$ every pair of clauses has at most $\lambda$ variables in common, and since we identify clauses with points of the BIBD, it is clear that any two points belong to at most $\lambda$ elements $\Phi(a_j)$ of $B$, i.e. blocks.

Now let the partial $(m,l,\lambda)$-BIBD be given with $m$ points $p_i$ and block set $B$. Obviously, condition (i) of the definition for an $l$-regular ${}^\lambda CNF$ is fulfilled with with $C_i = \Psi(p_i)$ since $\Psi(p_i) \in 2^B$ and $Var(F) = B$. (ii) is fulfilled because any two points $p_i, p_{i'}$ appear in at most $\lambda$ blocks, which means that the corresponding $C_i, C_{i'}$ share at most $\lambda$ blocks, i.e. variables.

(Note: The correspondence outlined above becomes apparent, when both structures are expressed in more modern language, i.e. as $l$-uniform hypergraphs.)

The correspondence theorem guaranties that at least the external architecture of an instance $F \in {}^\lambda CNF^l$ coincides with that of a partial $(m,l,\lambda)$-block design. Also their incidence matrices coincide. In particular $F \in X\,{}^\lambda CNF^l$ correspond to a BIBD.

Of course, CNF in general have an additional logical structure which demands that clauses are conjuncted and variables within clauses are disjuncted. However, Boolean satisfiability problems like exact satisfiability (XSAT) can be completely expressed via structural properties of the formula. To see this we start with a definition of exact satisfiability

*Definition 4*
A CNF formula $F$ with variable set $Var(F)$ is said to be exactly satisfiable or x-satisfiable or $F \in XSAT$ iff there is a truth assignment to the variables such that each clause contains exactly one true variable.

Such a truth assignment evaluates the whole formula to "true", of course. This definition can be phrased in purely structural terms without reference to logical terms as follows:

*Theorem* (partition is XSAT)
A CNF formula $F$ is x-satisfiable iff there is a subset $L \subseteq Var(F)$ which induces a partition of the clause set $F$.

*Proof*: Let $L = \{a_{j_1}, a_{j_2}, ..., a_{j_\alpha}\}$ be a subset of $Var(F)$, such that the subsets $F_s := \Phi(a_{j_s})$ partition $F$. Therefore $F_s \cap F_{s'} = \varnothing$ for all $s, s' \in \mathbb{N}_\alpha$ and $F = \cup F_s$. Thus each $C_i$ contains exactly one $a \in L$. If we set every variable in $L$ to "true", all other variables to "false", we obviously x-satisfy $F$.

If vice versa the truth assignment is given we define $L$ as the subset consisting of all variables set to "true" and the $F_s$ as before. Since no more than one true variable are allowed to occur in any $C$, there can be no clause which appears in two different $F_s$. Thus again $F_s \cap F_{s'} = \varnothing$, and since each $C$ must belong to at least one $F_s$ we have a partition of $F$, induced by $L$.

We will call such a set $L$ an XSAT-solution or short "solution" in the following. The variables of a solution are pairwise independent in the sense that they have no clause in common, i.e. they do not belong to any common clause. In a $XLCNF^l$ for each variable there are $v = n - 1 - l(k-1) = (l-1)(k-l)(k-1)/l$ independent variables, see [6] for a proof. There is no simple way, however, to determine sets of pairwise independent variables. The complexity class of XSAT restricted to $XLCNF^l$ is undetermined, but definitely of subexponential order [3].

The structural property that represents XSAT of CNF formulas has a counterpart in BIBDs:

*Definition 5*
a.     Two blocks of a $(m,l,\lambda)$-BIBD are *parallel* iff they have no point in common.
b.     A set of pairwise parallel blocks which partitions the point set is called a *parallel class* or *resolution*.
c.     A $(m,l,\lambda)$-BIBD is called *resolvable* iff the block set can be partitioned into parallel classes.

The term parallel stems from the fact that some simple BIBDs represent certain geometries in which blocks represent lines. That parallel lines have no points in common is immediately clear.
With this definition one can deduce from the correspondence theorem and the partition theorem the following problem correspondence:

To find an XSAT-solution for an $l$-regular $X^\lambda CNF$ is equivalent to finding a parallel class in the corresponding $(m,l,\lambda)$-BIBD.

The even more stringent notion of *resolvability* in design theory, see part c. of definition 5, is a key feature in paradigmatic problems like Kirkmans 'schoolgirl' problem, the Social Golfer problem or Eulers Officers problem, see e.g. [7,8]. Transferred to Boolean satisfiability it is a much more ambitious notion than XSAT. We may call it *complete x-satisfiability*, meaning that there are several pairwise disjoint solutions which exhaust the variable set.

**Transfers.**
In the following we will apply the correspondence theorem to transfer knowledge from the field of BIBDs to Boolean formulas and vice versa. We mentioned already the application of formula (BD1) to $l$-regular Boolean formulas and state:

*Lemma 1*:
Any $l$-regular exactly $\lambda$-connected CNF formula is uniform, i.e. each clause has the same number $k$ of variables, given by formula (BD1).

In other words: regularity forces uniformity in exactly $\lambda$-connected CNF formulas and in particular in exactly linear CNF formulas (i.e. $\lambda = 1$), XLCNF[l]. This fact apparently was not proven before 2018, [3]. In the theory of block designs, however, it is known for decades. Via the correspondence theorem it can be readily transferred to $l$-regular exact linear Boolean formulas.

Nevertheless we will give a simple proof of (BD1) using the incidence matrix $g_{ij}$. Condition (iii) of definition 3 may be expressed as $\sum_{j=1}^{n} g_{rj} g_{sj} = (1-\delta_{rs})\lambda + \delta_{rs} \sum_{j=1}^{n} g_{rj}$. Summing both sides over $r$ and using $\sum_{r=1}^{m} g_{rj} = l$ (because of (ii)) gives after rearranging terms $(l-1)\sum_{j=1}^{n} g_{sj} = \lambda(m-1)$. Since this holds for any point $s$ and since $(m,l,\lambda)$ are given constants, the number of blocks in which a point appears is equal for all points, $\sum_{j=1}^{n} g_{sj} = k_s = k$.

(Note: Relation (BD2) is a simple consequence of counting the entries of the incidence matrix column by column ($nl$) versus row by row ($km$).)

Note that the proof and conditions (BD1,2) only hold for BIBD and thus for exact $\lambda$-connected $l$-regular CNF formulas. For $l$-regular LCNF ($\lambda \in \{0,1\}$) the more general equations $m = 1 + \bar{k}(l-1) + \bar{d}$ and $nl = m\bar{k}$ are valid, where $\bar{k}$ is the mean uniformity parameter $\bar{k} = \sum_{i=1}^{m} |C_i|/m$ and $\bar{d}$ the mean number of clauses <u>not</u> connected ($\lambda = 0$) to other clauses, see [6] for details.

There is an extensive literature on necessary and sufficient conditions for a block design to exist. All these can now be transferred to derive statements about the existence of $l$-regular $X^\lambda CNF$ instances, of course. We will give just one example. *Fishers inequality $n \geq m$* is a well known necessary condition for the existence of a $(m,l,\lambda)$-BIBD. Thus one can infer from (BD2) that also $k \geq l$ is a necessary condition. Via correspondence we can thus state

*Lemma 2*:
The class $X^\lambda CNF^l$ is empty for $l > k$.

In other words, there are no $l$-regular, $k$-uniform $\lambda$-connected CNF formulas for $k < l$. In particular no $XLCNF^l$ for $k < l$ exist. Lemma 2 is a generalization of this fact stated as observation 1 in [2].

Conditions for the allowed parameters in an $l$-regular $X^\lambda CNF$ / $(m,l,\lambda)$-BIBD can readily be derived from (BD1,2) or simple considerations. E.g., $m$ must be a multiple of $l$ for the existence of a solution/parallel class, $m = \alpha l, \alpha \in \mathbb{N}$. For complete x-satisfiability/resolvability $n$ must be a multiple of $\alpha$, and the proportionality factor is given by the uniformity parameter via (BD2) already: $n = km/l = k\alpha$. Necessary and sufficient conditions for the existence of a BIBD with a parallel class (and thus for an XLCNF instance to be x-satisfiable) are known for a few special cases. Numerous estimates on the minimum number of pairwise parallel blocks for BIBDs exist, see [9] and ref. therein. These estimates can be transferred to estimates for the minimum number of pairwise independent variables of $X^\lambda CNF$.

A Steiner triple system $STS(m)$ is defined as a $(m,3,1) - BIBD$. Many results have been formulated for Steiner triple systems only. E. g., necessary and sufficient conditions for the existence of a $STS(m)$ are either $m = 1 \mod(6)$ or $m = 3 \mod(6)$, see e.g. [10]. Via the correspondence to $XLCNF^3$ we can rephrase this fact as

*Lemma 3*
3-regular exact linear CNF formulas exist if and only if 3 divides either $k$ or $k-1$.

*Proof*: From (BD1) $m-1 = 2k$ which is either a multiple of 6 or equals 2 plus a multiple of 6, from the existence criterion for the corresponding $STS(m)$.

If one is interested in XSAT solutions of 3-regular XLCNF a further restriction arises from the necessary condition $m = 0 \mod(l)$ which leads via (BD1) to $k = 1 \mod(l)$ as a necessary condition for a solution to exist. Together with the existence conditions of lemma 3 for XLCNF[3] this proves:

*Lemma 4*:
Instances of XLCNF[3] with $k \neq 1 \mod(3)$ exist, but none of these is x-satisfiable.

Another well known fact for BIBDs is that $m = l \mod(l(l-1))$ is a necessary and sufficient condition for the existence of a resolvable $(m,l,1)$-BIBD for $l = 3$ and $l = 4$, see [11]. Note that $k = 1 \mod(l)$ automatically fulfills this condition. Thus via correspondence we can state

*Lemma 5*:
$\{F \in XLCNF^l : k = 1 \mod l\} \neq \emptyset$ (at least) for $l = 3$ and 4. And there are even completely x-satisfiable instances in these classes.

As a last example for the use of the correspondence between BIBDs and Boolean satisfiability we prove a new result for partial Steiner systems by means of known theorems on x-satisfiability in $l$-regular linear CNF. To this end we formulate the *resolution problem for balanced incomplete block designs* as follows:

*Definition 6*
*Resolution problem*: Does a $(m,l,\lambda) - BIBD$ have a resolution?

We ask for the complexity of this problem, following a suggestion of Pak Ching Li and Toulouse, [4]. They posed the question for Steiner triple systems, i.e. for $(m,3,1)$-BIBDs and were able to answer the quest for <u>partial</u> Steiner triple systems where pairs of points belong to either one or no triple. With the use of correspondence we are able to get rid of the triple condition:

*Theorem* (resolution problem in partial BIBDs)
The resolution problem restricted to partial $(m,l,1)$-BIBDs is NP-complete.

*Proof*: Theorem 4 of [2] states that XSAT when restricted to the class $LCNF_+^l$ is NP-complete for all $l \geq 2$ (note that we assume monotone formulas throughout this paper). Via correspondence theorem one may identify $LCNF_+^l$ formulas with $m$ clauses to partial $(m,l,1)$-BIBDs. As was shown before, determining exact satisfiability for an instance $F \in LCNF^l$ corresponds to finding a parallel class in the corresponding partial block design, i.e. a partial $(m,l,1)$-BIBD in this case. This completes the proof.

This is a generalization of the cited result for triple systems, $l = 3$, to arbitrary $l$.

**Conclusion.**
A well defined correspondence between (partial) balanced incomplete block designs (BIBD) and regular (exact) linear Boolean formulas in conjunctive normal form ($LCNF^l$) is exploited to transfer results from one field to the other. An interesting new result is given for the problem determining the complexity class of finding a parallel class in certain partial BIBDs. With the aid of known facts on the complexity of the corresponding XSAT problem restricted to $LCNF^l$ one can show that this problem is NP-complete.

**References**.
[1] T. Schmidt, *Computational complexity of SAT, XSAT and NAE-SAT for linear and mixed Horn CNF formulas*, Ph.D. thesis, Institut für Informatik, Univ. Köln, Germany (2010).
[2] S. Porschen; T. Schmidt, E. Speckenmeyer, A. Wotzlaw, *XSAT and NAE-SAT of linear CNF classes*, Discrete Appl. Math. 167 (2014) 1-14.
[3] B.R. Schuh, *Exact satisfiability* of *linear CNF formulas*, Discrete Appl. Math. 251 (2018) 1-4.
[4] Pak Ching Li and Michel Toulouse, *Some NP-Completeness Results on Partial Steiner Triple Systems and Parallel Classes,* Ars Combinatoria 80 (2006) 45-51.
[5] Alexander Rosa, *Combinatorial design theory – notes*,
https://kam.mff.cuni.cz/conferences/ATCAGC_2014/materialy/DesignTheoryNotes.pdf
[6] B.R. Schuh, *Sub-exponential complexity of regular linear CNF formulas,*
*https://arxiv.org/ftp/arxiv/papers/1812/1812.09110.pdf* .
[7]Peter van Beek (Ed.) Principles and Practice of Constraint programing – CP 2005, Springer-Verlag 2005.
[8] Tony Hürlimann, Puzzles and Games, A Mathematical Modeling Approach. Department of Informatics, University of Fribourg, 2015 ISBN 9781326430894.
[9] Maria Di Giovanni, Mario Gionfriddo, and Antoinette Tripodi, *Parallelism in Steiner systems*, The Art of Discrete and Applied Mathematics 1 (2018) 1-9.
[10] C.J. Colbourn and A. Rosa, Triple Systems, Clarendon Press, Oxford (1999).
[11] H. Hanani, D.K.Ray-Chaudhuri, R.M.Wilson, *On resolvable designs*, Discrete Mathematics 306 (2006) 876-885.